\documentclass[times]{my}

\begin{document}

\title{The Probabilistic Explanation of the Cohort Component Population Projection Method }
\shorttitle{The Probabilistic Explanation of the Cohort Component Population Projection Method}

\author{Mariia Nosova\affil{1}}
\abbrevauthor{M. Nosova}
\headabbrevauthor{Nosova, M}

\address{%
\affilnum{1}Tomsk University of Control Systems and Radioelectronics, Tomsk, Russia}

\correspdetails{nosovamgm@gmail.com}

\received{}
\revised{}
\accepted{}

\communicated{ }

\begin{abstract}

The probabilistic explanation of the direct and reverse cohort component population projection methods is presented. The main characteristics determining the probability distribution of the values for the direct and reverse component methods are found. 
\end{abstract}

\maketitle

\section{Introduction}

In connection with the increasing role of the demographic factor in socio-economic planning, promising estimates of the size and composition of the population are relevant. Mathematical modeling is useful in solving this problem. The development and use of various kinds of mathematical models serve both for analysis of the reproduction of the population as a whole, and for revealing the patterns of development of certain demographic processes. In the modeling, certain initial assumptions are made regarding the main components of the process (fertility, mortality, migration, etc.). On this basis, other characteristics of the population and its structure are calculated. 

A special place in mathematical modeling is taken by the age-shifting method (or method of components) developed by P.K. Whelpton  \cite{bibid1}. The calculation of sex and age structure of the population by the cohort component method was conducted by S.G. Strumilin, A.Y. Boyarsky, P.P. Shusherin, M.S. Bednyi, S. Shcherbov, V. Lutz, W. Sanderson, and the UN Population Commission, The Federal Service for State Statistics, Center for Demography and Human Ecology \cite{bibid2, bibid3}. 

The cohort component method is effective enough for short-term forecasts with horizontal planning for not more than 10-15 years. In this paper, the probabilistic explanation of the direct and reverse cohort component population projection methods is presented. 

\section{Cohort Component Population Projection Method} 
The direct cohort component population projection method is used to determine the estimates of the number \textit{N(x+$\tau$,t+$\tau$)} of a group of persons of age \textit{x+$\tau$} in the year \textit{t+$\tau$} provided that the number \textit{N(x,t)} is known, and, \textit{$\tau$} is the prediction step. The population is considered in aggregate, without division by gender. Let \textit{p(x,x+$\tau$)} denote the conditional probability of reaching age \textit{x+$\tau$ }by persons of age \textit{x}.

It is known \cite{bibid4} that
\[p\left(x,x+\tau\right)=P\left\{X>x+\frac{\tau}{X}>x\right\}=\frac{P\left\{X>x+\tau,X>x\right\}}{P\left\{X>x\right\}}=\frac{S\left(x+\tau\right)\ }{S\left(x\right)},\] 
where \textit{S}(\textit{x}) is the survival function \cite{bibid4}, which is the probability that a person will survive to the age of \textit{x}. For \textit{N}(\textit{x},\textit{t}) and \textit{p}(\textit{x},\textit{x}+$\tau$), the probability distribution of the values of \textit{N}(\textit{x}+$\tau$\textit{,t}+$\tau$) is determined by the Bernoulli scheme and is binomial
\begin{equation} \label{GrindEQ__1_} 
P\{N(x+\tau,t+\tau)=m\}=C^m_{N\left(x,t\right)}p(x,x+\tau)^m(1-p\left(x,x+\tau\right))^{N\left(x,t\right)-m}, 
\end{equation} 
where the number of combinations is ${\textrm{C}}^m_{N(x,t)}=\frac{N(x,t)!}{m!(N(x,t)-m)!}$. Bernoulli's scheme is a series of \textit{N(x,t)} experiments, these experiments are independent, in each experiment, the event may occur with probability \textit{p}(\textit{x},\textit{x}+$\tau$) or not occur with probability 1-\textit{p}(\textit{x},\textit{x}+$\tau$). In terms of demography, this means that by \textit{N(x,t)} we mean the population size age \textit{x }in year \textit{t}. The success is the event that a randomly chosen person \textit{x} will live to age \textit{x + $\tau$}. The probability of such an event is \textit{p}(\textit{x},\textit{x}+$\tau$), the probability of failure is 1-\textit{p}(\textit{x},\textit{x}+$\tau$), respectively. It is known that the mathematical expectation for the binomial distribution has the form
\begin{equation} \label{GrindEQ__2_} 
EN(x+\tau,t+\tau)=N(x,t)p(x,x+\tau)=N(x,t)\frac{S(x+\tau)}{S(x)}.                                        
\end{equation} 

 Denoting the estimate of the value of the quantity \textit{N(x+$\tau$,t+$\tau$)} by the same symbol, we rewrite \eqref{GrindEQ__2_} in the form
\begin{equation} \label{GrindEQ__3_} 
N(x+\tau,t+\tau)S(x)=N(x,t)S(x+\tau)+\epsilon_1,                                                
\end{equation} 
where $\epsilon_1$ is a random error with mathematical expectation $E\epsilon_1=0$.

Equality \eqref{GrindEQ__3_} is the main one for applying the component method. In particular, for the direct component method, it is written in the form
\begin{equation} \label{GrindEQ__4_} 
N(x+\tau,t+\tau)=N(x,t)\frac{S(x+\tau)}{S(x)}+\epsilon_2,                                       
\end{equation} 
where \textit{N(x,t)} is given, and \textit{N(x+$\tau$,t+$\tau$)} is the estimate of the population number of the demographic group of persons of age \textit{x+$\tau$} in the year \textit{t+$\tau$}, while ${\epsilon}_2$ is a random error with mathematical expectation $E\epsilon_2=0$.

If we replace the argument \textit{x }on \textit{x}-$\tau$ while \textit{t} on \textit{t}-$\tau$ we rewrite \eqref{GrindEQ__3_} in the form
\[N(x,t)S(x-\tau)=N(x-\tau,t-\tau)S(x)+\epsilon_3,\] 
where $\epsilon_3$ is a random error with mathematical expectation $M\epsilon_3=0$. From where we get
\begin{equation} \label{GrindEQ__5_} 
N(x-\tau,t-\tau)=N(x,t)\frac{S(x-\tau)}{S(x)}+\epsilon_4,                                                           
\end{equation} 
where \textit{N}(\textit{x},\textit{t}) is given, and \textit{N}(\textit{x}-$\tau$,\textit{t}-$\tau$)\textit{ }is the estimate of the population number of the demographic group of persons of age \textit{x}-$\tau$ in the year \textit{t}-$\tau$, ${\epsilon }_4$ is a random error with mathematical expectation $E\epsilon_4=0$. Equality \eqref{GrindEQ__5_} makes it possible to determine the estimate of the population size of a demographic group at past instants of time. We call this \textit{the reverse cohort component population projection method}. The estimate \textit{N}(\textit{x}-$\tau$,\textit{t}-$\tau$) requires additional investigation, which we will perform below.

It follows from equality \eqref{GrindEQ__1_} that the estimate of the number \textit{N}(\textit{x}+$\tau$,\textit{t}+$\tau$) obtained by direct component method has a variance
\[DN(x+\tau,t+\tau)=N(x,t)p(x,x+\tau)(1-p(x,x+\tau))=N(x,t)\frac{S(x+\tau)}{S(x)}\left(1-\frac{S(x+\tau)}{S(x)}\right),\] 
and the coefficient of variation \textit{V${}_{1}$}${}_{\ }$of this quantity is
\[V_1=\frac{\sqrt{DN(x+\tau,t+\tau)}}{EN(x+\tau,t+\tau)}=\frac{1}{\sqrt{N(x,t)}}\sqrt{\frac{S(x)}{S(x+\tau)}-1}.\] 

Let's define the limits of the values of the coefficient of variation \textit{V${}_{1}$}. Since the number of five-year age groups in the statistical data \cite{bibid5} of the Russian Federation is of the order of several million, the first factor $1/\sqrt{N(x,t)}$ is less than 10${}^{-3}$. Using statistical data on the dependence of the survival function on age and analyzing all possible values of the second factor for $\tau \in [1;45]$ years and $x\le 70$ years, we find that the second factor takes a maximum value of 12.578 for $\tau =45$  years. As a result, in this case, we get that the coefficient \textit{V${}_{1}$} has values less than 0.0126. Since the estimate \eqref{GrindEQ__4_}  has sufficiently high accuracy, the error $\epsilon_2\ $can be neglected. 

\section{Reverse Cohort Component Population Projection Method} 

Equation \eqref{GrindEQ__5_}, which determines the estimate of the number \textit{N}(\textit{x}-$\tau$,\textit{t}-$\tau$) in the 1. The reverse cohort component population projection method is obtained by the use of the direct cohort method, so it is necessary to find the characteristics of this estimate, in particular, its mathematical expectation and variance. For a given value of \textit{N}(\textit{x},\textit{t}) we find the probability distribution of the number \textit{N}(\textit{x}-$\tau$,\textit{t}-$\tau$) of a group of persons of the age \textit{x-}$\tau$ in the year \textit{t}-$\tau$.

By the Bayes Theorem, we can write     
\begin{equation} \label{GrindEQ__6_} 
P\{N(x-\tau,t-\tau)=m/N(x,t)=n\}=\frac{P\{N(x,t)=n/N(x-\tau,t-\tau)=m\}P\{N(x-\tau,t-\tau)=m\}}{\sum^\infty_{v=n}{P\{N(x,t)=n/N(x-\tau,t-\tau)=v\}P\{N(x-\tau,t-\tau)=v\}}}.             
\end{equation} 

Here, similarly to \eqref{GrindEQ__1_}
\begin{equation} \label{GrindEQ__7_} 
P\{N(x,t)=n/N(x-\tau,t-\tau)=m\}=C^n_mp(x-\tau,x)^n(1-p(x-\tau,x))^{m-n},                 
\end{equation} 
where $p(x-\tau ,x)=S(x)/S(x-\tau )$. A priori distribution $P\{N(x-\tau,t-\tau)=m\}$ will be assumed to be Poisson distribution with some parameter \textit{a}, whose value we will define below
\begin{equation} \label{GrindEQ__8_} 
P\{N(x-\tau,t-\tau)=m\}=\frac{a^m}{m!}e^{-a}.                                                  
\end{equation} 

We consider the sum
\[\psi(z)=\sum^\infty_{v=n}{z^vP\{N(x,t)=n/N(x-\tau,t}-\tau)=v\}P\{N(x-\tau,t-\tau)=v\}.\] 

For brevity, we denote $p(x,x+\tau)=p$. By \eqref{GrindEQ__7_} and \eqref{GrindEQ__8_}, the function $\psi$(\textit{z}) can be written in the form
\[\psi(z)=\sum^\infty_{v=n}{z^vC^n_vp^n(1-p)^{v-n}}\frac{a^v}{v!}e^{-a}=\]
\[=p^ne^{-a}\sum^\infty_{v=n}{z^v\frac{v!}{n!(v-n)!}(1-p)^{v-n}}\frac{a^v}{v!}=\frac{p^ne^{-a}}{n!}\sum^\infty_{v=n}{z^v\frac{1}{(v-n)!}(1-p)^{v-n}}a^v=\] 
\[=\frac{(apz)^n}{n!}e^{-a}\sum^\infty_{v=n}{\frac{z^{v-n}}{(v-n)!}}{\left[a(1-p)\right]}^{v-n}=\] 
\[=z^n\frac{(ap)^n}{n!}e^{-a}e^{az(1-p)}=z^n\frac{(ap)^n}{n!}{exp \left\{a\left[(1-p)z-1\right]\right\}\ }.\]

The generating function $\phi$(\textit{z}) of the distribution \eqref{GrindEQ__6_} has the form
\begin{equation} \label{GrindEQ__9_} 
\phi\left(z\right)=\sum^\infty_{m=n}{z^mP\left\{N\left(x-\tau,t-\tau\right)=\frac{m}{N\left(x,t\right)}=n\right\}}=\frac{\psi(z)}{\psi(1)}=z^n{exp \{\ }(z-1)a(1-p)\}.     
\end{equation} 

Thus, the distribution \eqref{GrindEQ__6_} is the convolution of the degenerate distribution of the deterministic quantity \textit{n }and the Poisson distribution with the parameter
\begin{equation} \label{GrindEQ__10_} 
\lambda=a(1-p)=a\{1-p(x-\tau,x)\}.                                       
\end{equation} 

Let us find the a posteriori mean value of the quantity \textit{N}(\textit{x}-$\tau$\textit{,t-}$\tau$). It is obviously possible to write down
\[EN(x-\tau,t-\tau)=n+a\{1-p(x-\tau,x)\}.\] 

Assuming that the a priori and a posteriori mean values coincide, we write equation
\[a=n+a\{1-p(x-\tau,x)\},\] 
from which we find the values of the parameter \textit{a} in the form
\begin{equation} \label{GrindEQ__11_} 
a=n/p(x-\tau,x)=n\frac{S(x-\tau)}{S(x)} .                                             
\end{equation} 

Thus, the distribution \eqref{GrindEQ__6_} is determined by the generating function \eqref{GrindEQ__9_} with \textit{a} parameter of the form \eqref{GrindEQ__11_}. Let us find the conditional mathematical expectation and variance of the quantity \textit{N}(\textit{x}-$\tau$,\textit{t}-$\tau$) under the condition that the equality \textit{n}=\textit{N}(\textit{x},\textit{t}) holds. It is obvious that equality holds
\begin{equation} \label{GrindEQ__12_} 
EN(x-\tau,t-\tau)=a=N(x,t)\frac{S(x-\tau)}{S(x)},                                       
\end{equation} 
which justifies the choice of the estimate in the form \eqref{GrindEQ__5_}.

Let us find the conditional variance of the value of the estimate \eqref{GrindEQ__5_} under the condition that \textit{n}=\textit{N}(\textit{x},\textit{t}). By virtue of \eqref{GrindEQ__9_}
\[DN(x-\tau,t-\tau)=a\{1-p(x-\tau,x)\}=N(x,t)\frac{S(x-\tau)}{S(x)}\left(1-\frac{S(x)}{S(x-\tau)}\right)=N(x,t)\left(\frac{S(x-\tau)}{S(x)}-1\right)\] 
and the coefficient of variation \textit{V${}_{2}$} amounts to
\[V_2=\frac{\sqrt{DN(x-\tau,t-\tau)}}{EN(x-\tau,t-\tau)}=\frac{1}{\sqrt{N(x,t)}}\sqrt{\frac{S(x)}{S(x-\tau)}\left(1-\frac{S(x)}{S(x-\tau)}\right)}.\] 

Here, similar to \textit{V${}_{1}$} we define the range of values {}{}of the coefficient of variation \textit{V${}_{2}$}. The first factor $1/\sqrt{N(x,t)}$ is less than $10^{-3}$. Similarly, \textit{V${}_{1}$}, having analyzed all possible values {}{}of the second factor for $\tau \in [1;45]$ years and $x\le 70$ years, we get that the second factor takes the maximum value of 0.489 for $\tau =45$. As a result, we have that the variation coefficient \textit{V${}_{2}$} is less than 10${}^{-3}$ for any values {}{}of \textit{$\tau$} and \textit{x}. We note that in the sense of the values {}{}of the coefficients of variation, the estimates obtained by reverse movement are an order of magnitude (10 times) more accurate than the estimates obtained by direct movement at the same forecasting horizon \textit{$\tau$.} Therefore, a random error $\epsilon_4$ here can also be neglected.

\section{Conclusion}
The direct and reverse cohort component population projection methods have probabilistic justification and can be used to calculate the number of age groups in the years between the census dates \cite{bibid6}. This method is a simple tool for demographic analysis and gives results quite adequate to reality.

\end{document}